\documentclass[preprint,superscriptaddress,amssymb,amsmath,aps,prl,nobibnotes]{revtex4-2}
\usepackage{soul}
\usepackage[pdftex]{graphicx}
\usepackage{amsmath}
\usepackage{color}
\usepackage{natbib}
\usepackage{xcolor}
\usepackage[english]{babel}
\usepackage[autostyle]{csquotes}
\usepackage{lineno}

\begin{document}

This manuscript has been authored in part by UT-Battelle, LLC, under contract DE-AC05-00OR22725 with the US DOE. The U.S. government retains and the publisher, by accepting the article for publication, acknowledges that the U.S. government retains a nonexclusive, paid-up, irrevocable, worldwide license to publish or reproduce the published form of this manuscript, or allow others to do so, for U.S. government purposes. DOE will provide public access to these results of federally sponsored research in accordance with the DOE Public Access Plan.

\pagebreak
%\linenumbers
\title{Orbital-selective oxygen holes in cuprate ladders beyond the Zhang–Rice paradigm} 

\author{Chengyun Hua$^\dag$}
\email{huac@ornl.gov}
\affiliation{Materials Science and Technology Division, Oak Ridge National Laboratory, Oak Ridge, Tennessee 37831, USA}
\author{Tianran Chen}\thanks{These authors contributed equally to this work.}
\affiliation{Department of Physics and Astronomy, the University of Tennessee, Knoxville, Tennessee 37996, USA}
\author{Isaac C. Ownby}
\affiliation{Department of Physics and Astronomy, the University of Tennessee, Knoxville, Tennessee 37996, USA}
\author{Colin L. Sarkis}
\affiliation{Neutron Scattering Division, Oak Ridge National Laboratory, Oak Ridge, Tennessee 37831, USA}
\author{Garrett Granroth} 
\affiliation{Neutron Scattering Division, Oak Ridge National Laboratory, Oak Ridge, Tennessee 37831, USA}
\author{Masaaki Matsuda} 
\affiliation{Neutron Scattering Division, Oak Ridge National Laboratory, Oak Ridge, Tennessee 37831, USA}
\author{Jiaqiang Yan}
\affiliation{Materials Science and Technology Division, Oak Ridge National Laboratory, Oak Ridge, Tennessee 37831, USA}
\author{Ho Nyung Lee }
\affiliation{Materials Science and Technology Division, Oak Ridge National Laboratory, Oak Ridge, Tennessee 37831, USA}
\author{Jeongkeun Song}
\affiliation{Materials Science and Technology Division, Oak Ridge National Laboratory, Oak Ridge, Tennessee 37831, USA}
\author{Yuya Shinohara}
\affiliation{Materials Science and Technology Division, Oak Ridge National Laboratory, Oak Ridge, Tennessee 37831, USA}
\author{Masatomo Uehara}
\affiliation{Department of Physics, Yokohama National University, Yokohama 240-8501, Japan}
\author{Jun Akimitsu}
\affiliation{Department of Engineering Science, The University of Electro-Communications, Chofugaoka 1-5-1, Chofu, 182-8585, Tokyo, Japan}
\author{Oleksandr Prokhnenko}
\affiliation{Helmholtz-Zentrum Berlin, BESSY II, Albert-Einstein-Straße 15, Berlin, 12489, Germany}
\author{Eugen Weschke}
\affiliation{Helmholtz-Zentrum Berlin, BESSY II, Albert-Einstein-Straße 15, Berlin, 12489, Germany}
\author{Takeshi Egami}
\affiliation{Materials Science and Technology Division, Oak Ridge National Laboratory, Oak Ridge, Tennessee 37831, USA}
\affiliation{Department of Physics and Astronomy, the University of Tennessee, Knoxville, Tennessee 37996, USA}
\affiliation{Department of Materials Science and Engineering, the University of Tennessee, Knoxville, Tennessee 37996, USA} 
\affiliation{Shull Wollan Center, Oak Ridge National Laboratory, Tennessee 37831, USA} 
\author{D. Alan Tennant}
\email{dtennant@utk.edu}
\affiliation{Department of Physics and Astronomy, the University of Tennessee, Knoxville, Tennessee 37996, USA}
\affiliation{Department of Materials Science and Engineering, the University of Tennessee, Knoxville, Tennessee 37996, USA} 
\affiliation{Shull Wollan Center, Oak Ridge National Laboratory, Tennessee 37831, USA}

% Include the date command, but leave its argument blank.

\date{\today}

%%%%%%%%%%%%%%%%% END OF PREAMBLE %%%%%%%%%%%%%%%%

% Make the title.

\begin{abstract}

The electronic structure of the spin-ladder cuprate Sr$_{14}$Cu$_{24}$O$_{41}$ challenges the presumed universality of the Zhang–Rice singlet (ZRS) framework and models based exclusively on Cu–O hybridized orbitals. Combining polarization-dependent resonant soft X-ray scattering at the O $K$-edge with inelastic neutron scattering, we show that doped holes in the Cu$_2$O$3$ ladders localize predominantly in planar non-bonding O $2p_z$ ($p_\pi$) orbitals of rung oxygen sites rather than forming conventional ZRS states. Polarization-resolved RSXS uniquely identifies this orbital assignment, while lattice and magnetic excitations reveal its coupled consequences, establishing a unified microscopic picture that excludes the conventional $\sigma$-bonded singlet. This oxygen-sublattice charge order produces an anomalous diagonal stretching phonon and explains the absence of incommensurate magnetic fluctuations and anomalous magnon splitting. These findings motivate a reassessment of hole pairing in ladder cuprates and the sufficiency of copper-centric models for cuprate superconductors.

\end{abstract}

\maketitle

\pagebreak 
 
The cuprate superconductors have long defied a unified microscopic picture, yet one feature has proven remarkably robust: the holes introduced by doping settle predominantly on oxygen rather than copper~\cite{kowalski_oxygen_2021}. Extensive studies have shown that the doped holes on the planes go primarily into the O $2p$ state rather than the Cu $3d$ states~\cite{nucker_experimental_1987, tranquada_x-ray-absorption_1987, chen_electronic_1991, gauquelin_atomic_2014}. Theoretical efforts often begin with the one-band Hubbard model, which captures strong electron–electron correlations and Cu–O hybridization leading to the formation of Zhang–Rice singlet (ZRS) ground states~\cite{zhang_effective_1988, sheshadri_connecting_2023}. While this model qualitatively reproduces major features of the cuprate phase diagram, it remains too simplified to fully describe the intricate interplay between superconductivity and charge order. A more comprehensive framework is provided by the three-band Hubbard model~\cite{emery_theory_1987, emery_mechanism_1988, chen_doping_2013, wang_magnon_2018}, which explicitly includes the Cu $3d_{x^2-y^2}$ and O $2p_{\sigma}$ orbitals. Central to both models is the broad consensus that doped holes primarily occupy $p_\sigma$ oxygen orbitals, with non-bonding $p_\pi$ oxygen orbitals considered secondary to the low-energy physics---a picture presumed to hold across the full range of cuprate geometries, from the two-dimensional CuO$_2$ planes to the quasi-one-dimensional two-leg ladders~\cite{scheie_cooper-pair_2025,PhysRevX.15.021049}. 

%Here, we present experimental evidence that fundamentally challenges this paradigm: we demonstrate that doped holes in a cuprate ladder are not accommodated by ZRSs, but are instead carried by an orbital-selective oxygen $p_\pi$ channel, which has not been observed up to now. A corresponding charge order state develops primarily on the oxygen sublattice, inducing a local lattice distortion while leaving the copper spin ladder remarkably intact. By demonstrating the central role of the O $2p_\pi$ orbitals, these findings overturn the conventional low-energy description of doped cuprate architectures and open a new frontier in our understanding of correlated oxides.

Here, we present experimental evidence that fundamentally challenges this paradigm: we demonstrate that doped holes in a cuprate ladder are not accommodated by ZRSs, but are instead carried by an orbital-selective oxygen $p_\pi$ channel that has not been observed previously. Determining the orbital character of doped holes in strongly correlated materials is inherently challenging because individual experimental probes access only limited projections of a highly entangled electronic state. As a result, microscopic interpretations often remain underconstrained when based on a single observable. In this work, we combine resonant soft X-ray scattering (RSXS) with inelastic neutron scattering (INS) measurements of lattice and magnetic excitations, providing complementary sensitivity to orbital symmetry, local bonding geometry, and magnetic exchange interactions. This multi-probe approach ties the $\pi$-orbital charge order to its consequences for structural excitations and magnetism within a single consistent picture.

Little attention has been paid to charge dynamics mediated by non-bonding oxygen orbitals in the high-$T_c$ community. However, accumulating experimental and theoretical evidence suggests that an oxygen-network-facilitated transport channel may be a fundamental, yet largely ignored, component of the high-$T_c$ puzzle. Early band-structure calculations~\cite{PhysRevLett.58.1028, birgeneau_magnetic_1988} and first-principles quantum chemistry simulations~\cite{guo_electronic_1988} indicated that in-plane non-bonding O $2p_{\pi}$ orbitals can lie energetically higher than both $2p_{\sigma}$ and out-of-plane $2p_{\pi}$ states, making them highly favorable for hole occupancy and electrical conduction. Furthermore, oxygen $K$-edge X-ray absorption spectroscopy (XAS) and photoemission studies~\cite{chen_doping_2013, PhysRevLett.103.087402} observed a distinct saturation in the doping-dependent intensity of the XAS hole features beyond the underdoped regime. This directly contradicts the ZRS picture, which predicts a continuous, monotonic increase with doping, sparking considerable debate over the breakdown of single- and three-band Hubbard approximations~\cite{wang_theory_2010, PhysRevLett.105.199701, PhysRevLett.105.199702}. More recently, comprehensive nuclear magnetic resonance (NMR) spin shift analyses across hole-doped cuprate families~\cite{bandur_two-carrier_2026} have revealed a universal two-carrier scaling law, suggesting that an extended band structure incorporating Cu $4s$ and O $2p$ orbitals is required to account for the multiple electronic spin components observed experimentally~\cite{PhysRevLett.87.047003}. Establishing direct experimental evidence for hole occupancy within these non-bonding O $2p_\pi$ orbitals would therefore be pivotal to achieving a complete description of cuprate electronic structure.

\begin{figure*}
\includegraphics[scale = 0.55]{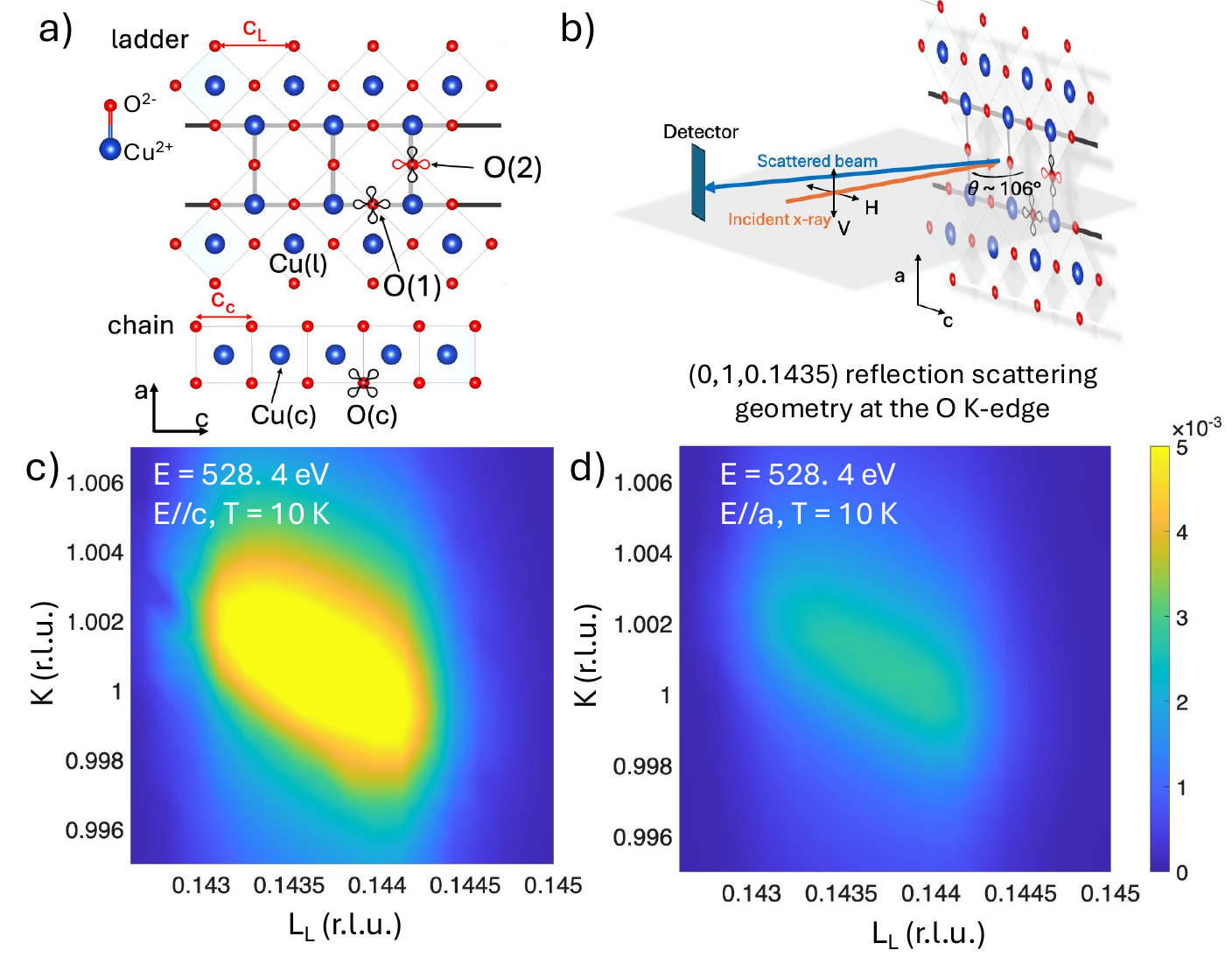}
\caption{Charge order revealed by RSXS. (a) Structure of the ladder (top) and chain (bottom) layers with a Cu-Cu distance denoted as $c_c$ and $c_L$, correspondingly. The ladders and chains are stacked alternatingly along the $b$-axis. Labeled oxygen sites include O(c) (chains), O(1) (ladder rails), and O(2) (ladder rungs). The red lobes indicate the O(2) $2p_z$ orbital analyzed in this study. The figure is adapted from Ref.~\cite{PhysRevB.62.14384}. (b) Horizontal scattering geometry showing the incident beam relative to the crystal lattice, with arrows indicating vertical and horizontal polarization. Intensity map of the charge order reflection at $(H, K, L_L)= (0, 1, 0.1435)$ in reciprocal space, measured at $E$ = 528.4 eV with (c) horizontal ($E \parallel c$) and (d) vertical ($E \parallel a$) polarization at 10 K. }
\label{fig:ChargeOrderPeak}
\end{figure*}

To isolate this subtle orbital character, spin-ladder compounds serve as a stripped-down model system. Despite their non-square layout, these systems share core features with their 2D counterparts---including a pseudogap, non-Fermi liquid behavior, and nearly identical coupling parameters---where the pairing of doped holes produces a superconducting state under pressure~\cite{dagotto_superconductivity_1992, dagotto_experiments_1999,vuletic_spin-ladder_2006}. In particular, the spin-ladder compound Sr$_{14}$Cu$_{24}$O$_{41}$ (SCO), extensively studied for its charge dynamics\cite{PhysRevB.54.15849, cox_low-temperature_1998,fukuda_periodic_2002,van_smaalen_comment_2003,etrillard_structural_2004, padma_symmetry-protected_2025,abbamonte_crystallization_2004, rusydi_strain_2008, PhysRevX.15.021049,padma_light-induced_2026,padma_symmetry-protected_2025}, is an ideal platform to probe the orbital character of charge carriers. Its crystal structure consists of alternating layers along the $b$-axis containing structurally and magnetically 1D CuO$_2$ chains and magnetically 2D, two-legged Cu$_2$O$_3$ spin ladders (see Fig.~\ref{fig:ChargeOrderPeak}a)~\cite{zimmermann_structural_2006, gotoh_structural_2003}. The intrinsic structural asymmetry of this Cu$_2$O$_3$ ladder network provides the exact orientational sensitivity needed to unambiguously pinpoint the oxygen orbitals hosting the doped holes.

Using polarization-dependent RSXS together with INS measurements of lattice and magnetic excitations, we show that holes within the Cu$_2$O$_3$ ladder subsystem do not form conventional ZRS states. Instead, they localize predominantly within the $\pi$-bonded O $2p_z$ orbitals of the rung oxygen sites (O(2) in Fig.~\ref{fig:ChargeOrderPeak}a). This orbital configuration leaves distinct and consistent fingerprints across the charge, lattice, and magnetic sectors, providing a unified microscopic picture of the doped ladder state and establishing direct experimental evidence for an oxygen $p_\pi$ hole channel in a cuprate.

RSXS measurements were performed on high-quality single crystals of Sr$_{14}$Cu$_{24}$O$_{41}$ at the UE46 beamline of BESSY II \cite{fink_resonant_2013}. Single crystals could be readily cleaved perpendicular to the $b$-axis, exposing a clean (010) surface suitable for a scattering geometry as depicted in Fig.~\ref{fig:ChargeOrderPeak}b. High-precision powder X-ray diffraction (XRD) measurement yields $c_C = 2.744 \pm 0.002 $\AA \ for the chain and $c_L = 3.932 \pm 0.002 $\AA \ for the ladder, corresponding to an incommensurate ratio of $c_C/c_L \approx 0.698 \pm 0.001$. For the purposes of reciprocal space indexing, a commensurate approximation of $c \approx 7c_L \approx 10 c_C$ defines the structural supercell~\cite{zimmermann_structural_2006,gotoh_structural_2003}. The Sr$_{14}$Cu$_{24}$O$_{41}$ crystal is the same one as used in Ref.~\cite{PhysRevLett.81.1702}.
Details about our samples and preparation and XRD analysis are provided in Supplementary Information (SI). 

At $T = 10$~K, a sharp resonant reflection is observed at $(H, K, L_L) = (0, 1, 0.1435)$, corresponding to $(0, 1, 1.005)$ in supercell notation and thus directly associated with the supercell periodicity. Here $L_L$ and $L$ are defined in reciprocal units of the ladder subunit and the supercell, respectively. The slight deviation from the integer $L = 1$ position reflects the aperiodic nature of the incommensurate crystal structure. The reflection is sharp in all three dimensions (see SI), with correlation lengths exceeding 100 unit cells, indicating long-range order. The asymmetry of the reflection may indicate a splitting, which is, however, not resolved here. Its intensity resonates at an incident photon energy of 528.4~eV, within the doped-hole feature of the O $K$-edge XAS. At this energy the reflection is accessed at a photon incidence angle of $106^\circ$---close to normal incidence (Fig.~\ref{fig:ChargeOrderPeak}b)---an ideal configuration for probing the relevant oxygen orbitals, since the X-ray polarization can readily be tuned along the $a$ and $c$ directions. The peak indeed exhibits a strong polarization dependence: it is prominently observed under horizontal polarization ($E \parallel c$) yet largely suppressed under vertical polarization ($E \parallel a$) (Figs.~\ref{fig:ChargeOrderPeak}c \& d).

This resonant reflection is fundamentally distinct from the superlattice feature reported in Refs.~\cite{abbamonte_crystallization_2004} and  \cite{rusydi_strain_2008}, which were characterized as a charge modulation confined strictly to the ladder and chain subsystem, respectively.  As shown in the following sections, the observed resonance displays a distinctive energy dependence, revealing charge ordering in both the chain and ladder subsystems and indicating a different charge order from earlier reports~\cite{PhysRevB.54.15849, cox_low-temperature_1998,fukuda_periodic_2002,van_smaalen_comment_2003,etrillard_structural_2004, padma_symmetry-protected_2025}

\begin{figure*}
\includegraphics[scale = 0.55]{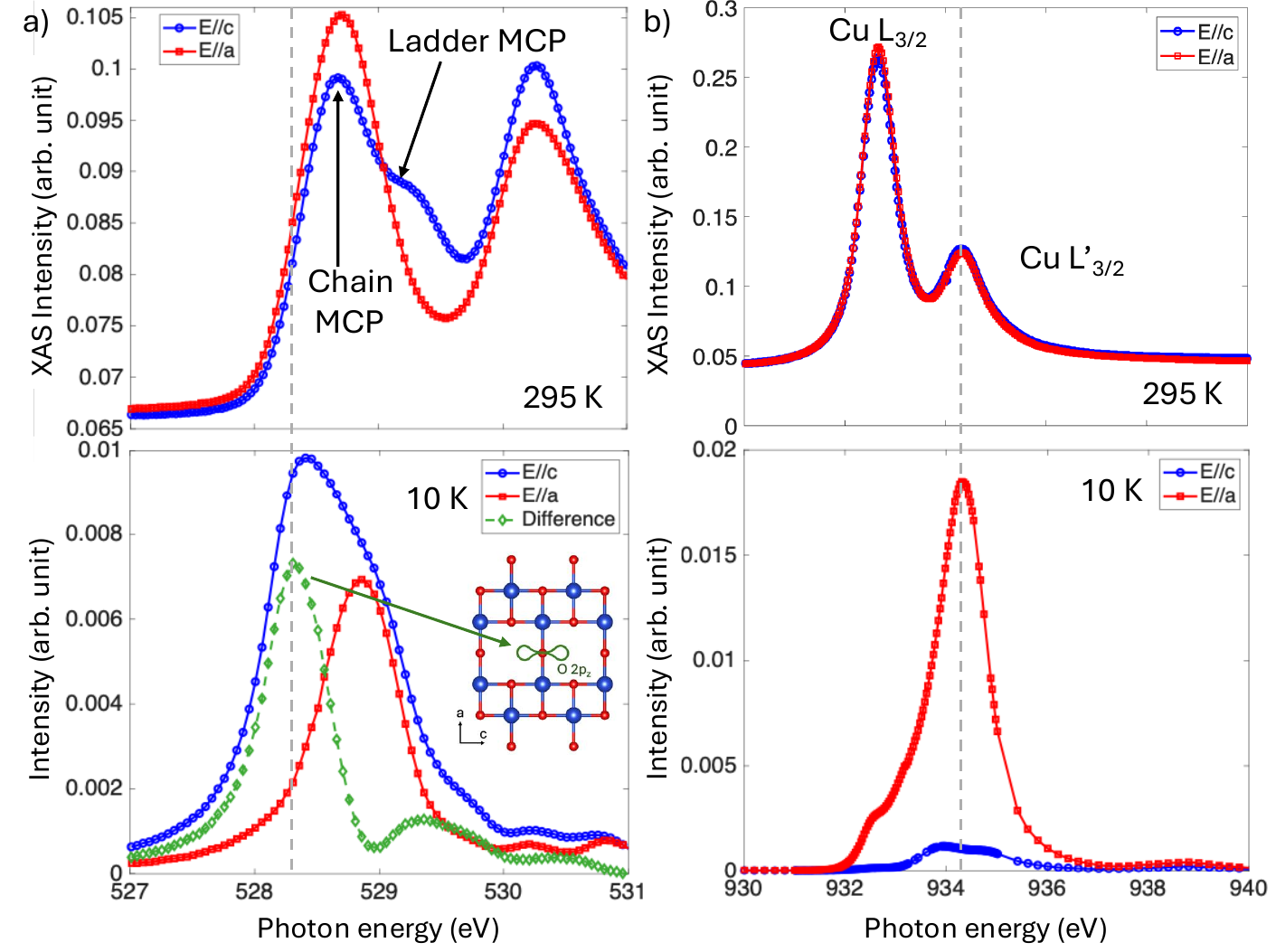}
\caption{ X-ray absorption spectra at room temperature (top) and the energy dependence of the $(0,1,0.1435)$ reflection at 10 K (bottom) of Sr$_{14}$Cu$_{24}$O$_{41}$ near (a) the O $K$ edge and (b) the Cu $L$ edge. Data are shown for horizontal ($E \parallel c$, blue circles) and vertical ($E \parallel a$, red squares) polarizations. The difference between the two polarizations at the O $K$-edge (green diamonds in (a), bottom) isolates a distinct resonant feature not resolved in the XAS spectra.}
\label{fig:EnergyScans}
\end{figure*}

Polarization-dependent XAS spectra near the oxygen $K$-edge are shown in the top panel of Fig.~\ref{fig:EnergyScans}a. The spectra were measured with high resolution (50 meV) by total electron-yield detection at room temperature in the geometry close to that of Fig.~\ref{fig:ChargeOrderPeak}b, yet at an incidence angle of exactly 90 degrees. The spectra are consistent with previous reports~\cite{PhysRevB.62.14384,huang_determination_2013,bugnet_real-space_2016}, revealing distinct mobile-carrier prepeaks (MCPs) associated with holes in the chain and ladder oxygen sites. The peak at 530.2 eV represents transition into the Cu $3d$ upper Hubbard band hybridized with O $2p$ states, while the MCP features near 528.7 eV correspond to transitions into O $2p$ doped-hole states in the conduction band. Specifically, the chain MCP, a peak at 528.7 eV, is assigned to the O(c) $2p_{x,z}$ orbitals (chain oxygen) and the ladder MCP, a shoulder near 529.3 eV, is assigned to the O(1) $2p_z$ orbital (rail oxygen)~\cite{PhysRevB.62.14384}. 

Tuning the incident photon energy to the MCP energies selectively enhances scattering from doped holes when they form an ordered state~\cite{abbamonte_structural_2002}. Consistent with this expectation, the energy dependence of the $(0,1,0.1435)$ reflection not only reveals resonant enhancement at the MCP energies but also exhibits distinct polarization-dependent behavior (see the bottom panel of Fig.~\ref{fig:EnergyScans}a). A previously unobserved resonance at 528.4~eV can be distinguished, and its polarization dependence identifies it with a resonant excitation into O(2) $2p_z$ orbitals.

The observation of this polarization-dependent resonance at 528.4~eV is significant for two primary reasons. First, it departs from the established phenomenology of charge modulation in planar cuprates, where the resonant energy typically coincides with, or lies in close proximity to, MCP features in XAS~\cite{PhysRevB.79.100502,abbamonte_crystallization_2004,abbamonte_spatially_2005}. Here, however, the resonance occurs at 528.4~eV---an energy located on the low-energy flank of the 528.9~eV XAS peak, where the spectral weight has already fallen below half of its maximum, and is therefore distinct from both the ladder (529.5~eV) and chain (528.9~eV) MCP energies. This displaced resonance shows that the underlying charge order arises from a distinct local electronic environment, rather than from a spatial modulation of the average doped-hole density. Owing to the low hole concentration in the ladders, this minority state remains largely invisible to site-averaged probes such as XAS but is readily detected through the phase-sensitive nature of resonant scattering~\cite{comin_resonant_2016}. 

Second, the resonance appears exclusively for $E \parallel c$, providing further insight into its microscopic origin. Given that the scattering geometry is close to normal incidence (sample $\theta \sim106^{\circ}$), the $E \parallel c$ configuration predominantly probes O~$1s \rightarrow$ O~$2p_z$ transitions; moreover, since both polarizations sample the same reflection $\mathbf{Q}$ from the same ordered domain with a similar geometric factor, the observed polarization contrast is intrinsic to the orbital channel rather than to scattering geometry. The ladder and chain subsystems contain only three crystallographically distinct oxygen sites---O(c), O(1), and O(2). With the chain O(c) feature at $\epsilon_{cx,z}=$ 528.9~eV and the O(1) $1s \rightarrow 2p_z$ transition already established at $\epsilon_{1z}=$ 529.5~eV~\cite{PhysRevB.62.14384}, the only remaining assignment for the third resonance, at $\epsilon_{2z}=$ 528.4~eV, is the O(2) $1s \rightarrow 2p_z$ transition into the non-bonding, in-plane $\pi$ orbital of the rung oxygen---a configuration that, being non-bonding, cannot form a ZRS. With the ordering wavevector fixing approximately one hole per ladder supercell~\cite{PhysRevB.62.14384,gotoh_structural_2003,huang_determination_2013}, this channel accounts for essentially all the ladder holes, leaving none for a $\sigma$-bonded singlet.

The hole origin of the reflection is also borne out at the copper $L$-edge. The Cu $L$-edge XAS spectrum has two primary features (Fig.~\ref{fig:EnergyScans}b, top): the main $L_{3/2}$ resonance and a high-energy satellite ($L'_{3/2}$), the latter arising from transitions into Cu $3d^{10}\underline{L}$ ligand-hole states~\cite{PhysRevB.79.100502}. The resonant profile of the reflection (Fig.~\ref{fig:EnergyScans}b, bottom) is dominated by these $3d^{10}\underline{L}$ states, consistent with the reflection also resonating at the oxygen MCP energies, which are intrinsically associated with $\sigma$-bonded oxygen orbitals. The intensity difference between the two polarizations at this edge arises primarily from the grazing scattering geometry required at the copper resonance: for $E \parallel c$, this configuration projects only a small component of the incident field onto the $c$-axis, reducing the transition matrix element. A detailed analysis of the Cu $L$-edge spectra and the associated geometric corrections is provided in SI.

\begin{figure*}
\includegraphics[scale = 0.4]{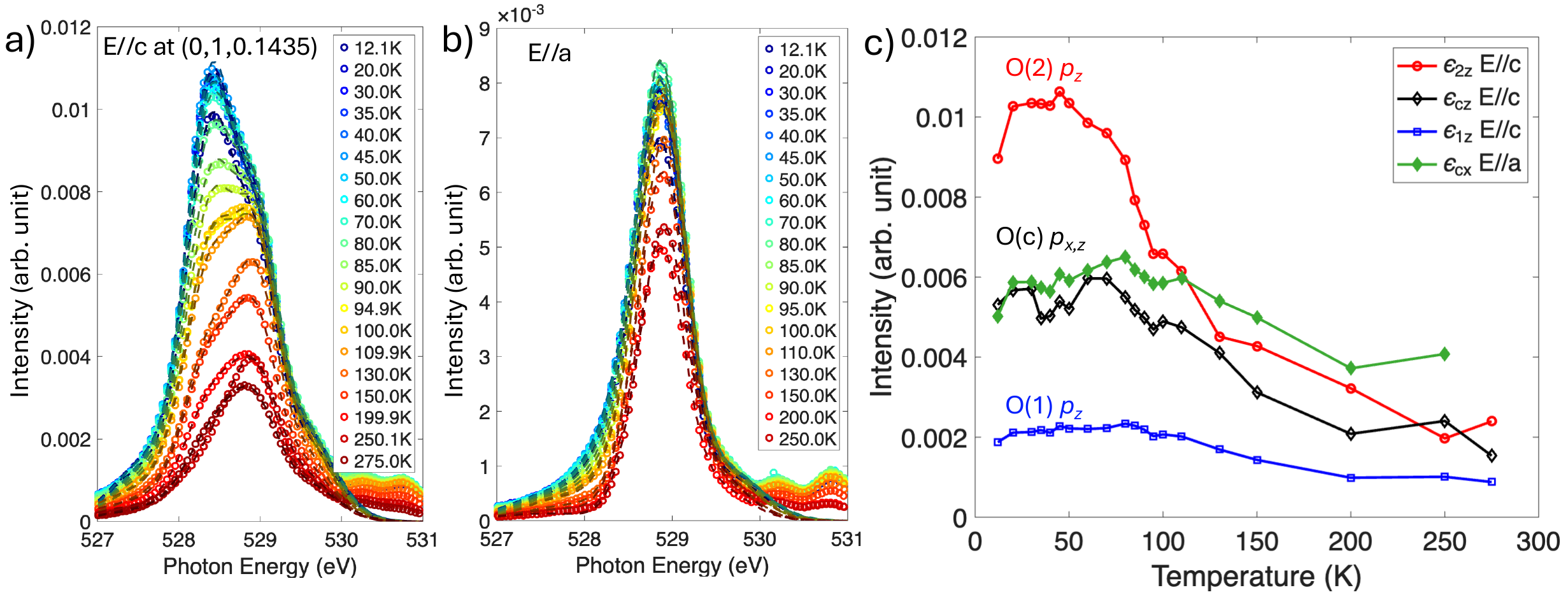}
\caption{ Temperature-dependent energy profiles of the $(0,1,0.1435)$ reflection measured from 12 K to 275 K using (a) horizontal polarization ($E \parallel c$) and (b) vertical polarization ($E \parallel a$). Experimental data are shown as symbols, and Gaussian fits are shown as dashed lines. (c) Temperature evolution of the resonant intensity at the fitted peak energies: $\epsilon_{2z}$ ($=$ 528.4~eV; associated with O(2) $2p_z$ orbitals, open circles),  $\epsilon_{cz}$ ($=$ 528.9~eV; associated with O(c) $2p$ orbitals, open diamonds), and $\epsilon_{1z}$ ($=$  529.5~eV; associated with O(1) $2p_z$ orbitals, open squares) for $E \parallel c$, and $\epsilon_{cx}$ ($=$ 528.9~eV; associated with O(c) $2p$ orbitals, solid diamonds) for $E \parallel a$.}
\label{fig:Tdep}
\end{figure*}

The temperature evolution of the $(0,1,0.1435)$ reflection reveals a further distinction of the newly observed charge order. Figure \ref{fig:Tdep} presents the energy-dependent intensity at fixed $\mathbf{Q}$, measured from 12 K to 275 K for both horizontal ($E \parallel c$) and vertical ($E \parallel a$) polarizations (Figs. \ref{fig:Tdep}a \& b). To extract the individual resonant contributions for $E \parallel c$, the spectra were decomposed following the observation of three distinct peaks in the resonance profile at 10 K (see Fig.~\ref{fig:EnergyScans}a, bottom; a detailed spectroscopic analysis is provided in SI). A three-Gaussian fit with peak positions fixed to the 10 K values ($\epsilon_{2z}=$ 528.4, $\epsilon_{cz}=$ 528.9, and $\epsilon_{1z}=$ 529.5 eV) is used for $E \parallel c$, whereas a single Gaussian at $\epsilon_{cx}=$ 528.9 eV suffices for $E \parallel a$ across the entire temperature range. The extracted temperature dependence of the peak intensities (Fig.~\ref{fig:Tdep}c) reveals a hierarchical melting of the charge order: the spectral weight of the $\epsilon_{2z}$ resonance begins to decrease at $\sim$50~K and reaches half of its low-temperature value by $\sim$130~K, while the remaining resonant peaks remain nearly constant up to $\sim$80~K before gradually diminishing. This bifurcated temperature dependence provides compelling evidence that the electronic origin of the $\epsilon_{2z}$ resonance is distinct from the other features, reinforcing its assignment to a unique O(2) $2p_z$ orbital occupancy.

Higher-order harmonic and near-zone-center satellite peaks---specifically at $(0, 1, 0.43)$ and $(0, 1, \pm 0.007)$ in $L_L$ unit---were observed. While these higher-order reflections carry important implications for the spatial extension and dimensionality of the hole-localization wavepacket, a detailed discussion is beyond the scope of this work, particularly as they appear in scattering geometries less favorable for polarization analysis or are inaccessible at the O $K$-edge resonance. Their observation nonetheless confirms that the reflection arises from a genuinely ordered state rather than an isolated resonance anomaly.

\begin{figure*}[!t]
\includegraphics[scale = 0.55]{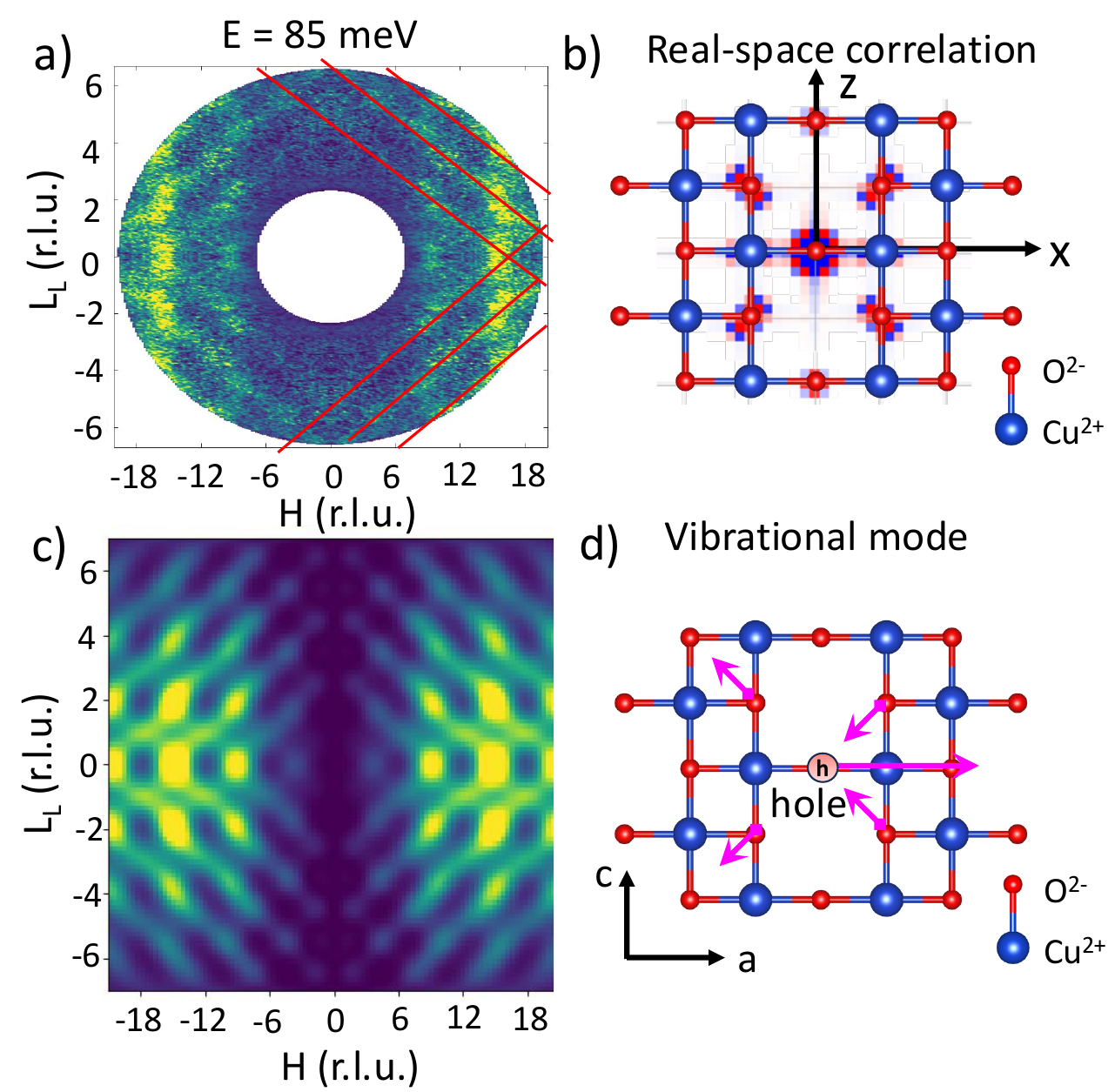}
\caption{Unconventional optical phonon mode. (a) Measured INS spectrum of Sr$_{2.5}$Ca$_{11.5}$Cu$_{24}$O$_{41}$ in the $(H0L_L)$ plane at 10 K, integrated over 85--90 meV; $L_L$ is in reciprocal lattice units of the ladder subunit. Beyond the vertical and weak horizontal stripes, the spectrum shows anomalous diagonal stripes (red lines) along $[H/3 \pm L_L]$, which are notable because the crystal lacks global symmetry along the $\hat{a} \pm \hat{c}$ directions. (b) Real part of the two-dimensional inverse Fourier transform of the stripe pattern in (a); red and blue denote positive and negative amplitudes, respectively. The mode center at the origin maps onto the O(2) rung oxygen site of the Cu$_2$O$_3$ ladder ($x$ and $z$ denote the crystallographic $a$ and $c$ axes).
(c) Simulated dynamic structure factor of the identified localized mode, reproducing the stripe pattern in (a). (d)  Schematic of the mode overlaid on the ladder; magenta arrows mark atomic displacements---the central O(2) rung oxygen vibrates along the rung ($a$-axis) direction, while the four neighboring O(1) leg oxygens move diagonally. The accompanying horizontal displacements of the two flanking Cu atoms (Cu--O(2)--Cu) are omitted for clarity; see SI for the complete diagram.}
\label{fig:Phonons}
\end{figure*}

A direct structural fingerprint of this orbital-specific hole occupancy is a local lattice distortion, evidenced by the emergence of an anomalous phonon mode within the cuprate ladder system. Our recent INS measurements using the Wide-Angular Range Chopper Spectrometer (ARCS) at the Spallation Neutron Source (SNS) have unveiled a localized optical phonon mode---induced by the localized oxygen holes---within the ladder layers of Sr$_{2.5}$Ca$_{11.5}$Cu$_{24}$O$_{41}$, a doped variant of the parent compound with a higher ladder hole concentration. The Sr$_{2.5}$Ca$_{11.5}$Cu$_{24}$O$_{41}$ crystal is the same one as used in Ref.~\cite{scheie_cooper-pair_2025}.

\textcolor{black}{The RSXS and INS data come from different members of the Sr$_{14-x}$Ca$_x$Cu$_{24}$O$_{41}$ series ($x = 0$ and $11.5$), but probe the same physics. The two compounds are chosen by purpose: $x = 0$ has the largest charge-order periodicity, placing its small ordering wavevector within reach of the O $K$-edge for the polarization-resolved RSXS orbital assignment, while the higher ladder doping at $x = 11.5$ amplifies the lattice and magnetic signatures probed by INS. Isovalent Ca$^{2+}$/Sr$^{2+}$ substitution leaves the total hole count fixed while redistributing holes from the chains into the ladder, raising the ladder concentration without changing the local rung coordination~\cite{gotoh_structural_2003,huang_determination_2013}. The orbital character of a ladder hole is fixed by the local Cu–O rung geometry, which is common to both compositions; Ca substitution tunes only the number of holes transferred into the ladders.  We therefore expect the same $\pi$-orbital occupancy in the $x = 11.5$ ladder, consistent with the amplified INS signatures reported below.}

The INS spectra reveal an unconventional vibrational feature at $85\text{ meV}$ in the $HL$-plane (Fig. \ref{fig:Phonons}a). This spectrum is characterized by three distinct features: vertical stripes along the $L$-direction, weak horizontal features along the $H$-direction, and anomalous diagonal stripes along the $[H/3 \pm L]$ directions. The emergence of these diagonal features is particularly striking, as the crystal structure lacks global symmetry along the $\hat{a} \pm \hat{c}$ directions.

To identify its microscopic origin, we computed the two-dimensional inverse Fourier transform of the measured pattern. The resulting real-space correlation map (the real part of the inverse Fourier transform of the measured intensity; Fig. \ref{fig:Phonons}b) comprises a central circular spot surrounded by eight anisotropic satellites. The pair at $(0, \pm c_L)$ arises from the horizontal stripes and is identified as a magnon contribution. The pair at $(\pm a/6, 0)$, displaced from the origin by one $\mathrm{Cu\text{-}O}$ bond length $d_\mathrm{Cu\text{-}O}$, arises from the vertical stripes and indicates a Cu--O bond stretching motion along the rung. The four satellites at $(\pm a/6, \pm c_L/2)$, at a distance $\sqrt{2}d_\mathrm{Cu\text{-}O}$ from the origin, arise from the diagonal stripes (red lines in Fig.~\ref{fig:Phonons}a) and correspond to stretching of diagonal O--O bonds. The real-space coordinates place the mode center unambiguously at a rung oxygen O(2) site, with the full satellite geometry matching the local bonding environment of the Cu$_2$O$_3$ ladder.

This diagonal coupling is itself orbital-selective. Stretching of the O(2)–O(1) diagonal bonds requires direct O $2p$–O $2p$ overlap along $\hat{a}\pm\hat{c}$, which only the O(2) $2p_z$ orbital---oriented along $c$ and projecting toward the rail oxygen---can provide; a hole in the $\sigma$ orbital ($2p_x$, directed along the rung toward Cu) would stiffen the Cu–O(2) bond alone and leave the diagonal O–O overlap untouched, and therefore cannot generate the observed diagonal satellites. The mode's energy points the same way: lying at $\sim$85~meV, roughly 10~meV above every other optical branch in the system (see SI for the full phonon model and dispersion), it requires the local O–O force constants to be substantially enhanced. Both the orientation and the local stiffening fingerprint the proposed orbital configuration: occupation of the O(2) $2p_z$ orbital modifies the local bonding character, increasing the effective O–O force constant and breaking the translational equivalence of the rung oxygens, so hole-bearing sites pin a vibrational mode above the phonon continuum while unoccupied sites retain the bulk spectrum. This phonon pattern is essentially temperature independent between 10 and 250 K (see SI), in contrast to the RSXS charge order, which partially melts above $\sim$50~K. The local orbital character of the hole---and its effect on the O--O force constant---is therefore more robust than the long-range charge order state.

\begin{figure*}
\includegraphics[scale = 0.35]{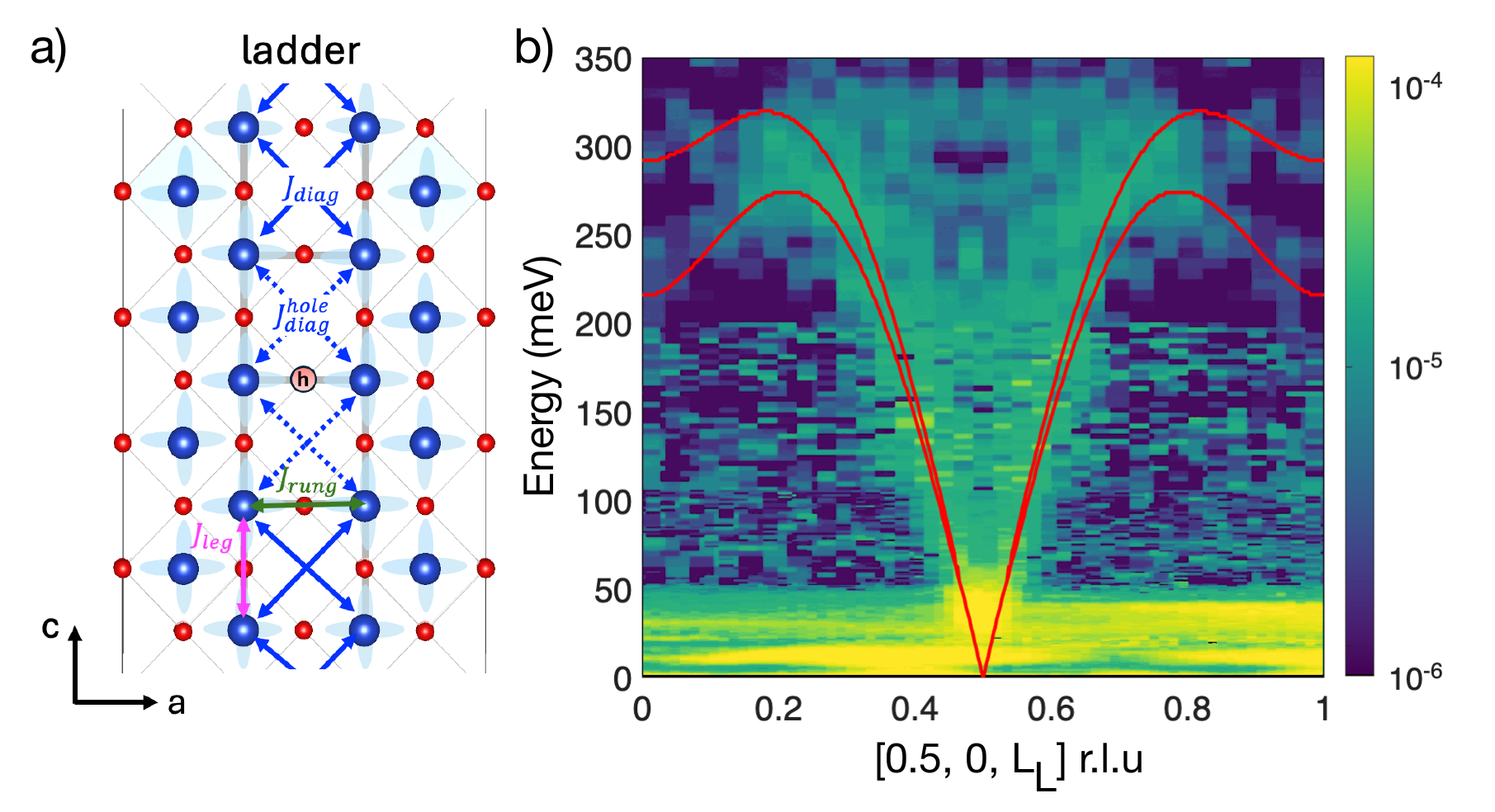}
\caption{Magnetic excitations and anomalous magnon dispersion in the ladder. (a) Heisenberg model of the ladder with leg ($J_{\mathrm{leg}}$) and rung ($J_{\mathrm{rung}}$) exchange and diagonal exchange $J_{\mathrm{diag}}$; solid blue arrows mark the normal diagonal exchange ($J_{\mathrm{diag}}^{\mathrm{normal}}$) and dashed blue arrows the suppressed exchange ($J_{\mathrm{diag}}^{\mathrm{hole}} \approx 0$) on plaquettes bridged by a hole-bearing O(2) rung oxygen (h symbol). (b) INS intensity maps ($E_i = 350$~meV) along $(1/2,0,L_L)$, showing a single magnon branch that splits into two away from the zone center; red lines are the binary-$J_{\mathrm{diag}}$ Heisenberg calculation. The intensity map are reproduced with permission from Ref.~\cite{scheie_cooper-pair_2025}. }
\label{fig:Spins}
\end{figure*}

%%%%%%%%%%%

Beyond its structural implications, the orbital-selective occupancy of holes has profound consequences for the magnetic response of the ladder subsystem. The $\pi$-orbital hole configuration identified here provides a natural explanation for the puzzling absence of magnetic signatures of hole doping in Sr$_{2.5}$Ca$_{11.5}$Cu$_{24}$O$_{41}$ \cite{scheie_cooper-pair_2025}. In canonical CuO$_2$ planes, doped holes reside in $\sigma$-bonded orbitals and form ZRSs, disrupting antiferromagnetic correlations and producing incommensurate magnetic fluctuations. By contrast, INS measurements reveal no such fluctuations in the ladders, but instead a robust spin gap despite substantial hole doping. Similarly, a separate study using resonant inelastic X-ray scattering (RIXS) at the Cu $L$-edge also shows that the magnetic response of hole carriers is strongly suppressed in Sr$_{14}$Cu$_{24}$O$_{41}$ \cite{PhysRevX.15.021049}. 

Previous interpretations based on single-band Hubbard models attributed this behavior to strong evidence of Cooper pair formation. Our results point to a different mechanism. If holes predominantly occupy $\pi$-bonded O(2) $2p_z$ orbitals, they are effectively decoupled from the $S=1/2$ Cu spins that govern the magnetic excitations. These “hidden” holes remain largely invisible to the spin sector, preserving an undoped-like antiferromagnetic response even at finite doping.

Nevertheless, orbital-selective hole occupancy leaves a clear fingerprint elsewhere in the magnon spectrum. High-energy INS~\cite{scheie_cooper-pair_2025} shows a single dispersive magnon branch near the zone center that splits into two as $|L_L - 1/2|$ grows, reaching $\sim$220 and $\sim$300~meV at the zone boundary along $(1/2,0,L_L)$ (Fig.~\ref{fig:Spins}b). A particularly important feature is the remarkably low spectral weight between the upper and lower branches. This observation places a strong constraint on the underlying exchange interactions. If the relevant exchange parameter varied continuously from rung to rung, one would generally expect a substantial distribution of magnon energies and corresponding intensity filling the region between the two branches. Instead, the spectrum remains sharply bifurcated, indicating that the ladder samples two discrete magnetic environments rather than a continuum of exchange values.

This behavior arises naturally within the orbital-selective hole picture. We modeled the ladder using a Heisenberg Hamiltonian with leg ($J_{\mathrm{leg}}$), rung ($J_{\mathrm{rung}}$), and effective diagonal ($J_{\mathrm{diag}}$) exchange, the last capturing the diagonal exchange characteristic of cuprate ladders~\cite{aligia_calculation_2018}. \textcolor{black}{This model is intended only as a heuristic illustration of how the observed two-branch dispersion could arise, not as a quantitative fit to the data.} A hole in the O(2) $2p_z$ orbital disrupts the diagonal exchange on its own plaquette, driving $J_{\mathrm{diag}}$ to zero there while leaving $J_{\mathrm{leg}}$ and $J_{\mathrm{rung}}$---which do not share that bridge---unchanged. A binary $J_{\mathrm{diag}}$---equal to its normal value on hole-free plaquettes and $\approx$ 0 on hole-bearing ones---reproduces the measured two-branch dispersion, including the zone-boundary energies and the merging at $L_L = 1/2$ (red lines, Fig.~\ref{fig:Spins}b); the upper branch arises from plaquettes with intact diagonal exchange and the lower from those where it is suppressed. The hole density sets only the relative weight of the two branches, not their positions, so the result is insensitive to its precise value (see SI for the supercell model and fit).Requiring only that localized holes suppress $J_{\mathrm{diag}}$ at O(2) rung sites, this qualitative agreement provides corroborating magnetic evidence for the orbital-selective hole localization independently inferred from RSXS and the anomalous phonon mode.

%%%%%%%%%%%%%
Taken together, these results provide evidence for $\pi$-orbital hole occupancy on the rung oxygen from three independent angles: the polarization-resolved orbital assignment from RSXS, the diagonal O–O stretching mode from the INS phonons, and the selective suppression of $J_{\mathrm{diag}}$ from the INS magnons. Notably, this interpretation was guided by the experimental data rather than by adopting a specific microscopic model a priori. The orbital assignment, the diagonal stretching motion, and the modulation of magnetic exchange were each inferred from correlated signatures across the spectroscopic, vibrational, and magnetic measurements before being incorporated into a unified physical picture.

More broadly, this work illustrates how combining complementary measurements can constrain microscopic interpretations of strongly correlated materials. By requiring consistency between charge, lattice, and spin observables, multi-modal experimental approaches can reduce ambiguities that often arise when individual probes are considered in isolation. In cuprates and related systems, where competing models frequently account for subsets of the available data, such integration provides an important framework for identifying electronic and structural degrees of freedom that may otherwise remain hidden.

\textcolor{black}{Corroboration from an entirely independent experimental axis comes from recent time-resolved measurements on this compound, which reported light-induced spectral weight near 528.4~eV in the O $K$-edge XAS and, in the same work, a transient collective excitation in time-resolved RIXS dispersing from a wavevector close to the charge order reflection reported here \cite{padma_light-induced_2026}. While those transient changes were attributed to chain-to-ladder hole transfer, our results suggest an alternative interpretation: light shuttles holes between the bonded and non-bonding O $2p$ orbitals.}

\textcolor{black}{Two fundamental questions arise from our findings: (1) what drives the preferential hole occupancy of the non-bonding $p_{\pi}$-orbital on the rung oxygens, and (2) what determines the resonance energy of this $p_\pi$ hole state?}

The preference for the $p_{\pi}$ state can be understood through a combination of structural pinning and local orbital relaxation. Our recent structural analysis of the ladder subunit revealed a modulated, out-of-plane distortion of the chain oxygens, whose periodicity and amplitude vary rigidly with Ca substitution~\cite{Chen_Lattice-fingerprint_2026}. This distortion projects a screened Coulomb potential along the rungs, creating periodically spaced potential minima that structurally pin charge carriers. On an electronic level, the dynamic Hubbard model~\cite{PhysRevB.90.184515} demonstrates that the local orbital relaxation triggered by electron removal drastically lifts the O $p_{\pi}$ band toward the Fermi energy. This massive relaxation energy shifts the energetic balance, overcoming the conventional hybridized $\sigma$-orbital preference and rendering the planar, non-bonding O $p_{\pi}$ orbitals highly favorable for hole accommodation.

\textcolor{black}{Addressing the second question requires a more complete description of the ladder electronic structure. The three-band Hubbard model retains only the O 2$p_\sigma$ orbitals that hybridize with Cu 3$d_{x^2-y^2}$; within this framework, all unoccupied O 2$p$ weight is tied to the $\sigma$-bonding network, and any O $K$-edge resonance must appear at the energies of the corresponding Zhang-Rice-like states---\emph{i.e.} at the ladder and chain MCP energies. The observed resonance at $\epsilon_{2z} =$ 528.4~eV, displaced from both, therefore has no natural counterpart in this model. The non-bonding $p_\pi$ orbitals, by contrast, are decoupled from the strong Cu--O $\sigma$ hybridization, so a hole occupying them experiences a different local potential and negligible ligand-field splitting from the Cu $d$ manifold, shifting its core-level excitation away from the MCP energies.  Capturing this behavior therefore requires a theoretical framework that extends beyond the conventional three-band Hubbard model by explicitly incorporating the non-bonding O $p_{\pi}$ states.}

For decades, high-temperature superconductivity has been viewed through a copper-centric lens, with single-band or three-band Hubbard models treating oxygen primarily as a passive ligand. By demonstrating that holes can selectively occupy a $\pi$-bonded O(2) $2p_z$ manifold, we uncover an alternative electronic channel that remains largely decoupled from conventional magnetic exchange pathways and bulk spectroscopic averages. This orbital-selective hole state, stabilized by a self-consistent lattice distortion, indicates that the oxygen network itself possesses an intrinsic electronic degree of freedom capable of driving emergent phases. In Sr$_{14}$Cu$_{24}$O$_{41}$, this mechanism manifests as a lattice-locked charge order, while the pronounced pressure dependence observed in doped ladder compounds may reflect a reorganization of oxygen $\pi$-orbital occupancy, calling for a reassessment of how pressure-induced electronic changes in these materials are interpreted. More broadly, if an analogous active $\pi$-oxygen network exists across the cuprate family, it could provide a microscopic route for connecting oxygen orbital physics to non-Fermi-liquid transport, competing orders in the pseudogap regime, and ultimately unconventional superconductivity, motivating a broader reconsideration of the role of oxygen throughout the cuprate phase diagram.

\section*{Acknowledgements}
We thank J\"{u}rgen Haase, Jorge Hirsch, Steve Johnston, Cristian Batista, and Elbio Dagotto for insightful discussions. The work of C.H., Y.S., C.S., G.G., M.M., J.Y., H.N.L., J.S., and T.E. was supported by U.S. Department of Energy (DOE), Office of Science, Office of Basic Energy Science (BES), Materials Sciences and Engineering Division. The work of T.C., I.C.O., and D.A.T. was supported by the University of Tennessee Materials Research Science \& Engineering Center - The Center for Advanced Materials and Manufacturing - supported by the National Science Foundation under DMR No. 2309083. 

RSXS Measurements were carried out at the XUV diffractometer (UE46 PGM-1) at the BESSY II electron storage ring operated by the Helmholtz-Zentrum Berlin für Materialien und Energie. A portion of this research used the resources at SNS, supported by DOE, BES, Scientific User Facilities Division. The beamtime was allocated to ARCS on Proposal No. IPTS32738.

\bibliographystyle{unsrt}
\bibliography{MyRef.bib}

\end{document}